\documentclass[journal,comsoc,onecolumn]{IEEEtran}

\usepackage[T1]{fontenc}% optional T1 font encoding

\usepackage{ifpdf}
\usepackage{cite}
\usepackage[pdftex]{graphicx}
\usepackage{amsmath}
\usepackage[cmintegrals]{newtxmath}
\usepackage{bm}
\usepackage{algorithm}
\usepackage{algorithmic}

\usepackage{array}

\ifCLASSOPTIONcompsoc
  \usepackage[caption=false,font=normalsize,labelfont=sf,textfont=sf]{subfig}
\else
  \usepackage[caption=false,font=footnotesize]{subfig}
\fi

\usepackage{enumerate}

\usepackage{dblfloatfix}
\ifCLASSOPTIONcaptionsoff
  \usepackage[nomarkers]{endfloat}
 \let\MYoriglatexcaption\caption
 \renewcommand{\caption}[2][\relax]{\MYoriglatexcaption[#2]{#2}}
\fi

\usepackage{url}

\usepackage{mleftright}
\mleftright %redefines all \left and \right to behave like \mleft and \mright

\newtheorem{theorem}{Theorem}

\newcommand{\vect}[1]{\mathbf{#1}} % vector
\newcommand{\const}[1]{\mathsf{#1}} % instead of Euler as IEEE doesn't know Euler font
\newcommand{\set}[1]{\mathcal{#1}}

\newcommand{\Prob}{\textnormal{Pr}}

\newcommand{\Prv}[1]{\textnormal{Pr}\mleft[#1\mright]}

\newcommand{\dd}{\mathop{}\!\mathrm{d}}
\newcommand{\ee}{\mathrm{e}}
\newcommand{\I}{\mathop{}\!\const{I}} %indicator function
\newcommand{\Qf}[1]{\ifstrequal{#1}{}{\mathcal{Q}}{\mathcal{Q}\mleft(#1\mright)}}

\newcommand{\LL}{\mathcal{L}}

\newcommand{\erf}[1]{\operatorname{erf}\mleft(#1\mright)}
\newcommand{\IG}{\mathrm{IG}}

\begin{document}
%
% paper title
% Titles are generally capitalized except for words such as a, an, and, as,
% at, but, by, for, in, nor, of, on, or, the, to and up, which are usually
% not capitalized unless they are the first or last word of the title.
% Linebreaks \\ can be used within to get better formatting as desired.
% Do not put math or special symbols in the title.
\title{The Crossover-Distance for ISI-Correcting Decoding of Convolutional Codes
  in Diffusion-Based Molecular Communications}
%
%
% author names and IEEE memberships
% note positions of commas and nonbreaking spaces ( ~ ) LaTeX will not break
% a structure at a ~ so this keeps an author's name from being broken across
% two lines.
% use \thanks{} to gain access to the first footnote area
% a separate \thanks must be used for each paragraph as LaTeX2e's \thanks
% was not built to handle multiple paragraphs
%

\author{Hui Li, Qingchao Li% <-this % stops a space
  \thanks{Hui Li and Qingchao Li are with the Department
of Electronic Engineering and Information Science, University of Science and
Technology of China, Hefei, Anhui, 230027, China (email:{mythlee}@ustc.edu.cn).
This work was supported by the National Science Foundation of China under Grant
No. 61471335.}% <-this % stops a space
}

% The paper headers
%\markboth{Journal of \LaTeX\ Class Files,~Vol.~14, No.~8, August~2015}%
%{Shell \MakeLowercase{\textit{et al.}}: Bare Demo of IEEEtran.cls for IEEE Communications Society Journals}
% The only time the second header will appear is for the odd numbered pages
% after the title page when using the twoside option.
% 
% *** Note that you probably will NOT want to include the author's ***
% *** name in the headers of peer review papers.                   ***
% You can use \ifCLASSOPTIONpeerreview for conditional compilation here if
% you desire.

% If you want to put a publisher's ID mark on the page you can do it like
% this:
%\IEEEpubid{0000--0000/00\$00.00~\copyright~2015 IEEE}
% Remember, if you use this you must call \IEEEpubidadjcol in the second
% column for its text to clear the IEEEpubid mark.

% use for special paper notices
%\IEEEspecialpapernotice{(Invited Paper)}

% make the title area
\maketitle

% As a general rule, do not put math, special symbols or citations
% in the abstract or keywords.
\begin{abstract}
  In diffusion based molecular communication, the intersymbol interference (ISI)
  is an important reason for system performance degradation, which is caused by
  the random movement, out-of-order arrival and indistinguishability of the
  moleclues. In this paper, a new metric called crossover distance is introduced
  to measure the distance between the received bit sequence and the probably
  tranmitted bit sequences. A new decoding scheme of conventional codes is
  proposed based on crossover distance, which can enhance the communication
  reliability significantly. The theoretic analysis indicates that the proposed
  decoding algorithm provides an approximately maximal likelihood estimation of
  the information bits. The numerical results show that compared with uncoded
  systems and some existing channel codes, the proposed convolutional codes
  offer good performance with same throughputs. 
\end{abstract}

% Note that keywords are not normally used for peerreview papers.

% For peer review papers, you can put extra information on the cover
% page as needed:
% \ifCLASSOPTIONpeerreview
% \begin{center} \bfseries EDICS Category: 3-BBND \end{center}
% \fi
%
% For peerreview papers, this IEEEtran command inserts a page break and
% creates the second title. It will be ignored for other modes.
\IEEEpeerreviewmaketitle

\section{Introduction}\label{sec:introduction}
Molecular communication (MC) is a promising information conveying paradigm for
nanoscale system, where due to the limitation of size and energy provided, the
traditional communication method using electromagnetic waves is difficult to
achieve between nano-devices. In molecular communication, the molecules, called
information molecules (particles) are released at the transmitter,
propagate through a fluid medium and are captured at the receiver. Various
modulation schemes are proposed to encoding the information on the properties of
molecules such as the number of released information particles, the
type/structure of particles or the time of released
\cite{Farsad_SurvyMC_2016CST}. The propagation of information molecules in fluid
medium follows Brownian motion in a random, out-of-order and indistinguishable
way, which make the reliable information transmission in molecular communication
is different from the conventional communication.

In order to achieve reliable information transmission, channel coding are usually
applied in the traditional communication systems to mitigate the effects of
noise and fading introduced by channel and electronic components. The early
works that directly use Hamming codes for on-off-keying in diffusion based MC
showed that when small number of molecules are used to encode each bit, uncoded
transmission outperforms coded transmission \cite{Leeson_FEC_2012NanoNet}. This
interesting issue occurs because the out-of-order arrival of information
particles at the receiver leads to severe intersymbol interference (ISI) which
dominates the error conditions. The performance of conventional convolutional
codes in concentration-encoded molecular communication is studied in
\cite{mahfuz_performanceCC_2013}, where ISI is also an important reason 
affecting bit error rate (BER) performance. In \cite{Lu_CompChannelCoding_2015IT},
conventional channel codes including hamming codes, Euclidean geometry LDPC, and
cyclic Reed-Muller codes are compared for diffusion-based molecular
communication systems.

Some new coding schemes tailored for MC channel are developed to cope with the
ISI problem. In \cite{ko_MoCoDF_2012Globalcomm}, molecular coding (MoCo)
distance function was proposed, which can measure the transition probability (or
distance) between the codewords affected by ISI. Then MoCo code is constructed
by maximizing the minimum pairwise MoCo distance. A new family of channel codes,
called ISI-free codes is proposed in \cite{shih_channelCodeMC_2013JSAC}, which
can eliminate the crossovers up to level-$l$ in one codeword and between
consecutive codewords. In \cite{kovacevic_zero-error_2014IT}, capacity-achieving
codes are constructed for a class of timing channel, called discrete-time
particle channel (DTPC), which can correct errors caused by random motion and
out-of-order arrival of particles.

In this paper, a new metric, called crossover distance is first proposed, which
measures the distance or crossover-transition probability between two binary
sequence. An algorithm with a time complexity $O(n^2)$ is provided to calculate
the crossover-distance efficiently, where $n$ is the length of the sequences.
Then based on crossover-distance we develop a decoding scheme for convolutional
codes to eliminate the errors caused by ISI in the diffusion-based MC channel.
The theoretic analysis indicates that the proposed
crossover-distance is optimal in the view of maximal likelihood under the
constraints on crossover-transition probability. The numerical results show that
the performance of crossover distance based decoding scheme of convolutional codes
outperforms conventional Hamming-distance based decoding scheme.
This shows that the proposed crossover distance reflects some essential
characteristics of the diffusion based MC channel.

The remainder of this paper is organized as follows. The system model is
described in Section \ref{sec:system-model}. In Section
\ref{sec:crossover-distance}, the proposed crossover distance is studied for
diffusion based molecular communications. The decoding scheme of convolutional
codes based on crossover-distance is provided in Section
\ref{sec:decoding-scheme}. Section \ref{sec:numerical-results} then gives the
numerical results and BER approximations of the proposed decoding scheme.
Finally, we conclude in Section \ref{sec:conclusion}.

\section{System Model}
\label{sec:system-model}

\subsection{Modulation and Diffusion Channel}
The communication system considered in this paper contians a pair of transmitter
and receiver. Binary Molecule Shift Keying (MoSK) modulation is used to
encode information bits \cite{kuran_modulationTech_2011ICC}, where two types of
distinguishable information particles are used to presented bit ``1'' and bit
``0'' respectively. Let $T_s$ be the fixed time interval of adjacent
transimition. At the beginning of each time slot, the transmitter emits a single
information particle where the type of the particle depends on the input being
bit ``1'' or bit ``0''.
% The transmitter emits one particle at the beginning of each
% time slot, $0, T_s, 2T_s,\cdots$ according to the bit to be sent, where $T_s$ is
% the fixed time interval of adjacent transmitted molecules.
Each information particle independently propagates in a 1-dimension diffusion
channel, arrives at the receiver disorderly and is captured by the receiver. The
receiver detect the type of each particle it captured and the particles are
removed from the system. In this paper we assume that the distance between the
transmitter and the receiver is $d$ and the particles diffuse from the
transmitter to the receiver with a drift velocity $v>0$. The random travel time
of information particles is the length of time between release from the
transmitter and the first hitting time which has an inverse Gaussian (IG)
distribution \cite{Chhikara1988} with the corresponding probability density
function (PDF)\footnote{The function $\I\{\cdot\}$ denotes the indicator
  function that takes on the values 1 or 0 depending on whether the statement
  holds true or not.}
\begin{align}
  \label{eq:IGpdf}
  f_{\IG}(t)=\sqrt{\frac{\lambda}{2\pi}} t^{-\frac{3}{2}}\exp\left( -\frac{\lambda(t-\mu)^2}{2\mu^2t} \right)\I\{t>0\}.
\end{align}
Here $\mu$ is the average travel time and $\lambda$ relates to the Brownian motion
parameter $D$:
\begin{align}
  \label{eq:IGparam}
  \mu=\frac{d}{v}, \qquad \lambda=\frac{d^2}{2D}, \qquad D=\frac{k_BT_{\mathrm{K}}}{6\pi\eta r}
\end{align}
where $k_B$ is the Boltzmann constant, $T_{\mathrm{K}}$ is the absolute
temperature, $\eta$ is the viscosity constant which depends on the liquid
type and its temperature, and $r$ is the  radius of molecules.

\begin{figure}
  \centering
  \includegraphics[width=0.5\textwidth]{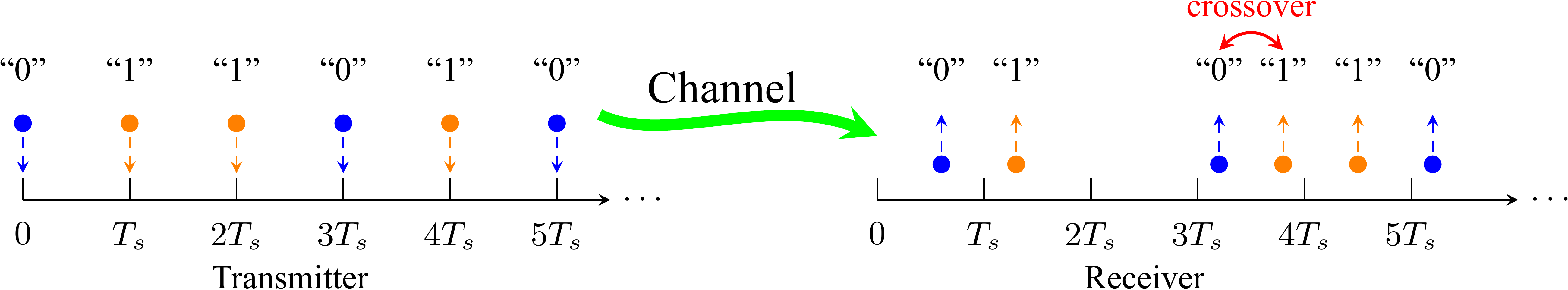}
  \caption{Binary MoSK modulation in duffsion channel}
  \label{fig:system-model}
\end{figure}

Fig. \ref{fig:system-model} illustrates the modulation and channel model in
this paper, where the transmitter encodes bits ``0,1,1,0,1,0'' into two types
(blue/orange) particles. Whereas at the receiver bits ``0,1,0,1,1,0'' are
decoded for there is a level-1 crossover happened at the third bit, which
introduces the inter-symbol interference.

\subsection{Crossover Transition Probability}

When the arriving order of the particles is different from their transmission
order, the order of the recovered bit sequences at the receiver is also
different from the order of the original one, similar to
\cite{shih_channelCodeMC_2013JSAC}, we call this phenomenon \emph{crossover}. A
level-$l$ crossover is defined to be the crossover of one bit delayed after $l$
bits, for example, in ``0\textbf{1}0001'' $\to$ ``000\textbf{1}01'',
there exists a level-$2$ crossover at the second bit. In this paper, the
background noise is not considered. The bit errors are only caused by the
intersymbol interference from the crossovers and this channel is denoted as ISI
MC channel.

In coding theory, the distance between codewords is an important concept, which
can be used to analyze the performance of codes, generate codewords and design
efficient decoding algorithms. In order to measure the distance of input and
output bit sequences of the above ISI channel, it is necessary first to
calculate the probability of a single level-$l$ crossover occurring, or
crossover transition probability $P_c(l)$. According to the given modulation and
channel model in this paper, suppose a level-$l$ crossover occurring at $i$th
bit, then $P_c(l)$ is
\begin{align}
  P_c(l)& =  \Prob [t_i>lT_s+t_{i+l}, t_i<(l+1)T_s+t_{i+l+1}, \nonumber \\
   & \qquad\quad t_i<(l+2)T_s+t_{i+l+2}, \cdots]   \label{eq:crossProb}
\end{align}
where $t_i$ is the travel time of $i$th particle and \eqref{eq:crossProb} means
that $i$th particle arrives after $(i+l)$th particle but before the particle at
$(i+l+1)$th, $(i+l+2)$th, $\cdots$.
For the sake of completeness, level-$0$ is used to indicate that no
crossover has occurred, and $P_c(0)$ is defined as
\begin{align}
  \label{eq:crossP0}
  P_c(0) & = 1-\sum_{l=1}^\infty P_c(l).
\end{align}
Given the probability of the first hitting time in formula
\eqref{eq:IGpdf}, and considering the independency of the particle's travel
time, ($t_{k}$, $k=i+l,i+l+1,\cdots)$, we have
\begin{align}
  \label{eq:PrCross}
  P_c(l) & = \int_0^{+\infty}\!\!\!f_{\IG}(u)\Prv{u>lT_s+t_{i+l}}\cdot \Prv{u<(l+1)T_s+t_{i+l+1}}\cdot\nonumber \\
         & \quad \Prv{u<(l+2)T_s+t_{i+l+2}}\cdots \dd u \\
         & =  \int_0^{+\infty}\!\!\!f_{\IG}(u)\Prv{t_{i+l}<u-lT_s}\prod_{k=l+1}^{\infty}\Prv{t_{i+k}>u-kT_s}\dd u \nonumber\\
         & = \int_0^{+\infty}\!\!\!f_{\IG}(u)F_{\IG}(u-lT_s)\prod_{k=l+1}^{\infty}[1-F_{\IG}(u-kT_s)] \dd u
\end{align}
where $F_{\IG}(\cdot)$ is the cumulative distribution function (cdf) of inverse
Gaussian distribution \cite{Chhikara1988}, defined on $(0,+\infty)$,
\begin{align}
  \label{eq:IGcdf}
  F_{\IG}(t)=\Phi\mleft[\sqrt{\frac{\lambda}{t}}\mleft(\frac{t}{\mu}-1\mright)\mright] + \ee^{\frac{2\lambda}{\mu}}\Phi\mleft[-\sqrt{\frac{\lambda}{t}}\mleft(\frac{t}{\mu}+1\mright) \mright].
\end{align}
In \eqref{eq:IGcdf}, $\Phi(z)=\frac{1}{2}\mleft(1+\erf{\frac{z}{\sqrt{2}}}\mright)$ is the cdf of a
standard Gaussian random variable $Z$.

\section{Crossover-Distance}
\label{sec:crossover-distance}
\subsection{Crossover Vector}
In order to measure the distance between input and output binary sequences of
the ISI channel, \emph{crossover vector} is first introduced, which can describe the
crossover status of each bit between the two binary sequences.

Let $\vect{x}=(x_1,x_2, \cdots, x_N)$ and $\vect{y}=(y_1, y_2, \cdots, y_N)$
be the input and output sequences of the ISI MC channel. It is assumed that the
number of ``0'' and the number of ``1'' in these two sequences are same, which
stems from the assumption that the bit errors are only caused by ISI. Define
crossover vector as $\vect{v}=(v_1,v_2,\cdots,v_N)$, where
\begin{align}
  \label{eq:cv_bit}
  v_i=\begin{cases}
    l & \textrm{if a level-$l$ crossover happened for bit $x_i$} \\
    0 & \textrm{if no crossover for bit $x_i$}
  \end{cases}.
\end{align}
For example vector $\vect{v}=(0,1,0,0,1,0)$ is the crossover vector from
$\vect{x}=(0,1,0,0,0,1)$ to $\vect{y}=(0,0,1,0,1,0)$, where two level-1
crossover are happened on the 2nd and 5th bits in $\vect{x}$.

Given the channel input and output sequences $\vect{x}$ and $\vect{y}$, the crossover vector
$\vect{v}$ can be constructed based on the one-to-one correspondence between the
bit $x_i$ in $\vect{x}$ and the bit $y_j$ in $\vect{y}$. Define $L$ as a
bijective function (mapping) on $\set{I}=\{1,2,\cdots,N\}$, satisfying 
\begin{align}
  \label{eq:LLdefine}
  L(i)=j \quad L^{-1}(j)=i \quad x_i=y_j \quad i,j\in \set{I}
\end{align}
where $\set{I}$ is the set of the index of each bit in $\vect{x}$
and $\vect{y}$. For any $k\in \set{I}$, the crossover level of $k$th bit $x_k$
under bijective mapping $L$, $l_{L,k}$ is
\begin{align}
  \label{eq:clevel_k}
  l_{L,k}=\mleft(\max_{{j\le L(k)}}L^{-1}(j)-k\mright)^+
\end{align}
where 
\begin{align}
  \label{eq:plusOpt}
  (n)^+=
  \begin{cases}
    n & n>0 \\
    0 & n\le 0.
  \end{cases}
\end{align}
The main idea in \eqref{eq:clevel_k} is to find the latest transmit bit which is
received before the current bit and calcuate the transmitting interval of these
two bits. Then the crossover vector can be constructed based
on~\eqref{eq:clevel_k} for each bit, $\vect{v}=(l_{L,1}, l_{L,2}, \cdots, l_{L,N})$.

There are many bijective functions which can be used to construct crossover
vector, let us denote the set of bijective function which can construct
crossover vector between $\vect{x}$ and $\vect{y}$ by $\LL(\vect{x}, \vect{y})$.
A special crossover vector with minimum summation and minimum maximization of
the crossover level, called \emph{minimum crossover vector} can be constructed as
follows: First generate the sorted ``0''-bit index vector and ``1''-bit index
vector for $\vect{x}$ and $\vect{y}$ respectively as $\vect{b}^{(x,0)},
\vect{b}^{(x,1)},\vect{b}^{(y,0)},\vect{b}^{(y,1)}$, where the $i$th element of
$\vect{b}^{(x,0)}$, $b^{(x,0)}_i$ is the index of the $i$th ``0''-bit in
$\vect{x}$, and $j$th element of $\vect{b}^{(x,1)}$, $b^{(x,1)}_j$ is the index
of the $j$th ``1''-bit in $\vect{x}$, the elements in $\vect{b}^{(y,0)}$ and
$\vect{b}^{(y,1)}$ are similar to these. Then construct a bijective function $L_0$
on $\set{I}$ as
\begin{align}
  \label{eq:L0}
  L_0(i)=
  \begin{cases}
    b^{(y,0)}_k & \mbox{if $i= b^{(x,0)}_k$} \\
    b^{(y,1)}_k & \mbox{if $i= b^{(x,1)}_k$}
  \end{cases}
\end{align}
Using the formula \eqref{eq:clevel_k}, the minimum crossover vector $\vect{v}_0$
then can be constructed and the minimum properties of $\vect{v}_0$ is proved
later in Theorem \ref{thm:crossvect}. To simplify the calculation, Algorithm
\ref{alg:crossvect} is presented as a construction method of minimum crossover
vector directly from $\vect{x}$, $\vect{y}$, $\vect{b}^{(x,0)}$,
$\vect{b}^{(x,1)}$, $\vect{b}^{(y,0)}$, $\vect{b}^{(y,1)}$, the computational
complexity of which is about $\mathrm{O}(N^2)$.
\begin{algorithm}
  \caption{Minimum crossover vector construction} \label{alg:crossvect}
  \begin{algorithmic}[1]
    \STATE given $\vect{x}$, $\vect{y}$, $\vect{b}^{(x,0)},
    \vect{b}^{(x,1)},\vect{b}^{(y,0)},\vect{b}^{(y,1)}$
    \FOR{$i=1$ to $N$}
    \IF{$x_i=0$}
    \STATE find $k_1$ such that $b_{k_1}^{(x,0)}=i$ and let $j=b^{(y,0)}_{k_1}$
    \STATE find $k_2$ where $k_2=\arg \max_{k} b^{(y,1)}_k$ and $b_{k_2}^{(y,1)}<j$
    \IF {$k_2$ does not exist or $b_{k_2}^{(x,1)}\le i$}
        \STATE $v_i=0$
    \ELSE 
    \STATE $v_i$ = $b_{k_2}^{(x,1)}-i$
    \ENDIF
    \ELSE 
    \STATE find $k_1$ where $b_{k_1}^{(x,1)}=i$ and let $j=b^{(y,1)}_{k_1}$
    \STATE find $k_2$ where $k_2=\arg \max_{k} b^{(y,0)}_k$ and $b_{k_2}^{(y,0)}<j$
    \IF {$k_2$ does not exist or $b_{k_2}^{(x,0)}\le i$}
        \STATE $v_i=0$
    \ELSE 
    \STATE $v_i$ = $b_{k_2}^{(x,0)}-i$
    \ENDIF
    \ENDIF
    \ENDFOR
    \STATE output $\vect{v}$.
  \end{algorithmic}
\end{algorithm}

An example of the minimum crossover vector generated by Algorthm
\ref{alg:crossvect} for given $\vect{x}$ and $\vect{y}$ is presented in Table
\ref{tab:exampleCrossover} .

\begin{table}[H]
  \centering
  \caption{An example of minimum crossover vector}
  \label{tab:exampleCrossover}
  \begin{tabular}{c|c||c|c || c | c}
    \hline
    $\vect{x}$ & 0\textbf{1}00\textbf{0}10 & $\vect{b}^{(x,0)}$ & $(1,3,4,5,7)$
    & $\vect{b}^{(x,1)}$ & $(2,6)$ \\  \hline
    $\vect{y}$ & 000\textbf{1}1\textbf{0}0 & $\vect{b}^{(y,0)}$ & $(1,2,3,6,7)$
    & $\vect{b}^{(y,1)}$ & $(4,5)$\\ \hline
    $\vect{v}$ &\multicolumn{5}{c}{$(0, \mathbf{2},0,0,\mathbf{1},0,0)$} \\ \hline
    $L_0$ & \multicolumn{5}{c}{$(1,3,4,5,7,2,6) \leftrightarrow (1,2,3,6,7,4,5)$} \\ \hline
  \end{tabular}
\end{table}

Let us denote the set of crossover vectors which can transform the channel input
$\vect{x}$ into the output $\vect{y}$ by  $\set{V}(\vect{x},\vect{y})$. 

\begin{theorem}\label{thm:crossvect}
  Let $\forall \vect{v} \in \set{V}(\vect{x},\vect{y})$ and $\vect{v}_0$ be the
  crossover vector constructed according bijective function $L_0$ in
  \eqref{eq:L0} and \eqref{eq:clevel_k}. Then $\vect{v}_0$ is minimum crossover
  vector, thus the following properties hold:
  \begin{enumerate}[a)]
  \item Define $S_{\vect{v}}$ as the sum of the elements in the crossover vector $\vect{v}$, then
      \begin{align}
        \label{eq:sum_min}
        S_{\vect{v_0}}\le S_{\vect{v}}
      \end{align}
    \item Define $M_{\vect{v}}$ as the maximum of the elements in the crossover
      vector $\vect{v}$, then
      \begin{align}
        \label{eq:max_min}
        M_{\vect{v_0}}\le M_{\vect{v}}
      \end{align}
  \end{enumerate}
\end{theorem}
\begin{IEEEproof}
  For any bijective function $L \in \LL(\vect{x},\vect{y})$ a crossover vector
  $\vect{v}$ can be generated according to \eqref{eq:clevel_k}. If $\exists i_1,
  i_2\in \set{I}$, $x_{i_1}=x_{i_2}$ and
  \begin{align}
    \label{eq:Li1i2}
    L(i_1)=j_1, \quad L(i_2)=j_2, \quad i_1<i_2, \ j_1>j_2,
  \end{align}
  a new bijective mapping $L_1(\cdot)$  can be constructed as
  \begin{align}
    \label{eq:Lnew}
    L_1(i)=\begin{cases}
      L(i) & \mbox{$i\neq i_1$ and $i\neq i_2$} \\
      j_2 & i=i_1 \\
      j_1 & i=i_2
    \end{cases},
  \end{align}
  and a new crossover vector $\vect{v}_1$ can also constructed from $L_1$. Next
  we will prove that $S_{\vect{v}_1}\le S_{\vect{v}}$ and $M_{\vect{v}_1}\le
  M_{\vect{v}}$.

  For any $k\in \set{I}$ the crossover level $l_{L,k}$ of $k$th bit $x_k$ under mapping
  $L$ can be calculated from \eqref{eq:clevel_k}, then we have
  \begin{enumerate}[i)]
  \item If $k\neq i_1$ and $k\neq i_2$, $L(k)=L_1(k)$, 
    \begin{itemize}
    \item if $L(k)<j_2$ or $L(k)>j_1$, $l_{L,k}=l_{L_1,k}$;
    \item if $j_2<L(k)<j_1$, $l_{L_1,k}\le l_{L,k}$; ``$<$'' if $i_2 =
      \max_{j<L(k)} L^{-1}(j)$ and $i_2-k>0$;
    \end{itemize}
    % \item if $k=i_1$, $L_1(i_1)=j_2<j_1=L(i_1)$, $l_{L_1,k}\le{L,k}$;
    \item If $k_1=i_1, k_2=i_2$, define $A$ and $B$ as
      \begin{align}
        \label{eq:LkmaxAB}
        A=\max_{j\le L(i_2)} L^{-1}(j) \qquad B=\max_{j\le L(i_1)} L_1^{-1}(j)
      \end{align}
      Since $L(i_2)<L(i_1)$ we have $A\le B$, $l_{L,i_1}=(B-i_1)^+$, $l_{L,i_2}=(A-i_2)^+$,
      $l_{L_1,i_1}=(A-i_1)^+$, and $l_{L_1,i_2}=(B-i_2)^+$. Then the crossover
      levels versus $A$ and $B$ is given in Table \ref{tab:ABtable}.
%       \begin{itemize}
%       \item if $A\le B < i_1 < i_2 $, 
%         $l_{L_1,i_1}=l_{L_1,i_2}=l_{L,i_1}=l_{L,i_2}=0$
%         $l_{L_1,i_1}+l_{L_1,i_2}=l_{L,i_1}+l_{L,i_2}$ and $\max \{l_{L_1,i_1},
%         l_{L_1,i_2}\}=\max \{l_{L,i_1}, l_{L,i_2}\} $
%       \item if $A<i_1 < B < i_2$, $l_{L_1,i_1}=l_{L_1,i_2}=0$,
%         $l_{L,i_1}=B-i_1$, $l_{L,i_2}=0$. 
%         % \begin{align*}
%         %   l_{L_1,i_1}+l_{L_1,i_2}&<l_{L,i_1}+l_{L,i_2} \\
%         %   \max \{l_{L_1,i_1}, l_{L_1,i_2}\}&<\max \{l_{L,i_1}, l_{L,i_2}\} 
%         % \end{align*}
%       \item if $i_1 < A \le B < i_2$, $l_{L_1,i_1}=A-i_1$, $l_{L_1,i_2}=0$,
%         $l_{L,i_1}=B-i_1$, $l_{L,i_2}=0$.  
%         % \begin{align*}
%         %   l_{L_1,i_1}+l_{L_1,i_2}&\le l_{L,i_1}+l_{L,i_2} \\
%         %   \max \{l_{L_1,i_1}, l_{L_1,i_2}\}&\le \max \{l_{L,i_1}, l_{L,i_2}\} 
%         % \end{align*}
%       \item if $i_1 < A  < i_2 < B$, $l_{L_1,i_1}=A-i_1$, $l_{L_1,i_2}=B-i_2$,
%         $l_{L,i_1}=B-i_1$, $l_{L,i_2}=0$.  
%         % \begin{align*}
%         %   l_{L_1,i_1}+l_{L_1,i_2}&\le l_{L,i_1}+l_{L,i_2} \\
%         %   \max \{l_{L_1,i_1}, l_{L_1,i_2}\}&\le \max \{l_{L,i_1}, l_{L,i_2}\} 
%         % \end{align*}
%       \item if $i_1 <  i_2 < A \le B$, $l_{L_1,i_1}=A-i_1$, $l_{L_1,i_2}=B-i_2$,
%         $l_{L,i_1}=B-i_1$, $l_{L,i_2}=A-i_2$.  
%         % \begin{align*}
%         %   l_{L_1,i_1}+l_{L_1,i_2}&= l_{L,i_1}+l_{L,i_2} \\
%         %   \max \{l_{L_1,i_1}, l_{L_1,i_2}\}&\le \max \{l_{L,i_1}, l_{L,i_2}\} 
%         % \end{align*}
%      \end{itemize}
      \begin{table}[H]
        \centering
        \caption{crossover level versus A and B}
        \label{tab:ABtable}
        \begin{tabular}{c|cc|cc}
          \hline 
          & $l_{L_1,i_1}$ & $l_{L_1,i_2}$ & $l_{L,i_1}$ & $l_{L,i_2}$ \\ \hline
          $A\le B <i_1<i_2$ &0 & 0 & 0 & 0\\  \hline
          $A<i_1 < B < i_2$ &0 & 0 & $B-i_1$ & 0 \\ \hline
          $i_1 < A \le B < i_2$ & $A-i_1$ & $0$ & $B-i_1$ & 0 \\ \hline
          $i_1 < A  < i_2 < B$ & $A-i_1$ & $B-i_2$ & $B-i_1$ & 0 \\ \hline
          $i_1 <  i_2 < A \le B$ & $A-i_1$ & $B-i_2$ & $B-i_1$ & $A-i_2$  \\ \hline
        \end{tabular}
      \end{table}
      It is easily to deduce
      \begin{align}
        l_{L_1,i_1}+l_{L_1,i_2}&\le l_{L,i_1}+l_{L,i_2} \\
        \max \{l_{L_1,i_1}, l_{L_1,i_2}\}&\le \max \{l_{L,i_1}, l_{L,i_2}\} 
      \end{align}
    \end{enumerate}
    Then we have
    \begin{align}
      \label{eq:SM}
      S_{\vect{v}_1}&\le S_{\vect{v}} \\
      M_{\vect{v}_1}&\le M_{\vect{v}}.
    \end{align}
    Iteratively constructing new mapping based on forumla \eqref{eq:Lnew}, the
    ordered bijective function can be achieved, where the index of $k$th ``0''
    bit in $\vect{x}$ corresponds to the index of $k$th ``0'' bit in $\vect{y}$,
    and the case of ``1'' bit is same. This is exactly the bijective function
    $L_0$ defined in \eqref{eq:L0}. Then the crossover vector $\vect{v}_0$ generated
    from $L_0$ satisfies \eqref{eq:sum_min} and \eqref{eq:max_min}. Thus Theorem
    \ref{thm:crossvect} is proved.
  \end{IEEEproof}

\subsection{Crossover Distance}
Given the crossover vector for the channel input sequence $\vect{x}$ and output
sequence $\vect{y}$, the distance between these two sequence can be defined
based on the probability of crossover occurring in the crossover vector. In this
paper, it is assumed that the crossovers occurred in one crossover vector are
independent. Then the probability that $\vect{x}$ is transformed to $\vect{y}$
according to $\vect{v}=(l_1,l_2,\cdots,l_N)$ can be calcuated as:
\begin{align}
  \label{eq:ProbV}
  P_{\vect{v}}=\Prv{\vect{v}:\vect{x}\to\vect{y}}=\prod_{k=1}^N P_c(l_k)
\end{align}
Then the crossover distance is defined as
\begin{align}
  \label{eq:crossdist}
  D_{\vect{v}}(\vect{x},\vect{y})& =-\log P_{\vect{v}} + N\log P_c(0) \\
                                 & = -\sum_{k=1}^N \log \frac{P_c(l_k)}{P_c(0)} \\
                                 & = \sum_{k=1}^N W_c(l_k) \label{eq:DvWeight}                                   
\end{align}
where $W_c(l_k)$ is the distance contribution of level-$l_k$ crossover in
$\vect{v}$, defined as
\begin{align}
  \label{eq:Wc}
  W_c(l) = \log P_c(0) - \log P_c(l)
\end{align}
To ensure that $D_{\vect{v}}(\vect{x},\vect{y})\ge0$, in this paper we assume
$P_c(0)> P_c(l) $ for $l\ge 1$. $D_{\vect{v}}(\vect{x},\vect{y})$ has the following
properties:
\begin{itemize}
\item When there is no crossover occurring, $\vect{v}=\vect{0}$,
  $\vect{y}=\vect{x}$, $D_{\vect{v}}(\vect{x},\vect{x})=0$.
\item If $D_{\vect{v_1}}(\vect{x}, \vect{y}) < D_{\vect{v_2}}(\vect{x},
  \vect{y})$,  $P_{\vect{v}_1} > P_{\vect{v}_2}$. In this
  paper the minimum crossover distance between $\vect{x}$ and $\vect{y}$ is
  estimated by using the minimum crossover vector $\vect{v_0}$ between
  $\vect{x}$ and $\vect{y}$, i.e. $D_{\vect{v}_0}(\vect{x},\vect{y})$. For sake
  of simplification, we denote $D(\vect{x},
  \vect{y})=D_{\vect{v}_0}(\vect{x},\vect{y})$.
\item $D(\vect{x_1}, \vect{y})<D(\vect{x_2},\vect{y})$ indicates that given the
  channel output $\vect{y}$, $\vect{x_1}$ is sent with a higher probability
  than $\vect{x_2}$.
\item $D(\vect{x}, \vect{y_1})<D(\vect{x},\vect{y_2})$ indicates that given the
  channel input $\vect{x}$, $\vect{y_1}$ is received with a higher probability
  at receiver than $\vect{y_2}$.
\end{itemize}

In summary, the proposed crossover distance can be used to measure the distance
between the codewords in channel codes for the ISI molecular channel in the view
of maximum likelihood (ML) probability.

\section{New Decoding Scheme of Convolutional Codes}\label{sec:decoding-scheme}
The crossover distance proposed in Section
\ref{sec:crossover-distance} can be applied to develop various channel codec
scheme for the ISI molecular channel. In this paper, we use the crossover
distance as a metric in the decoding of the convolutional codes and a new viterbi
decoding scheme is proposed which shows good performance compared to
conventional channel codes such as hamming codes and hamming distance based convolutional codes.

\subsection{Encoding scheme}
\label{sec:coding}

\begin{figure}
  \centering
  \includegraphics[width=0.5\textwidth]{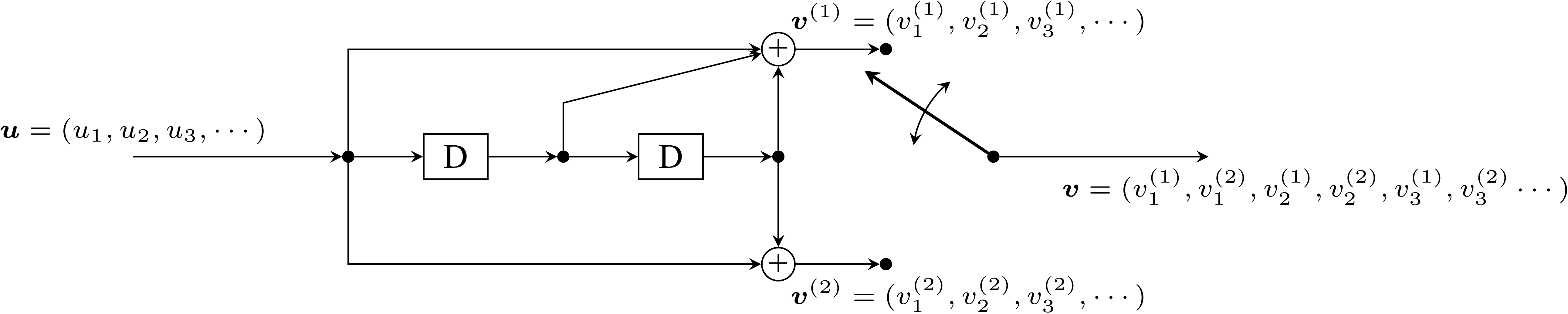}
  \caption{Encoding structure of convolutional codes}
  \label{fig:encoding}
\end{figure}

\begin{figure}
  \centering
  \includegraphics[width=0.5\textwidth]{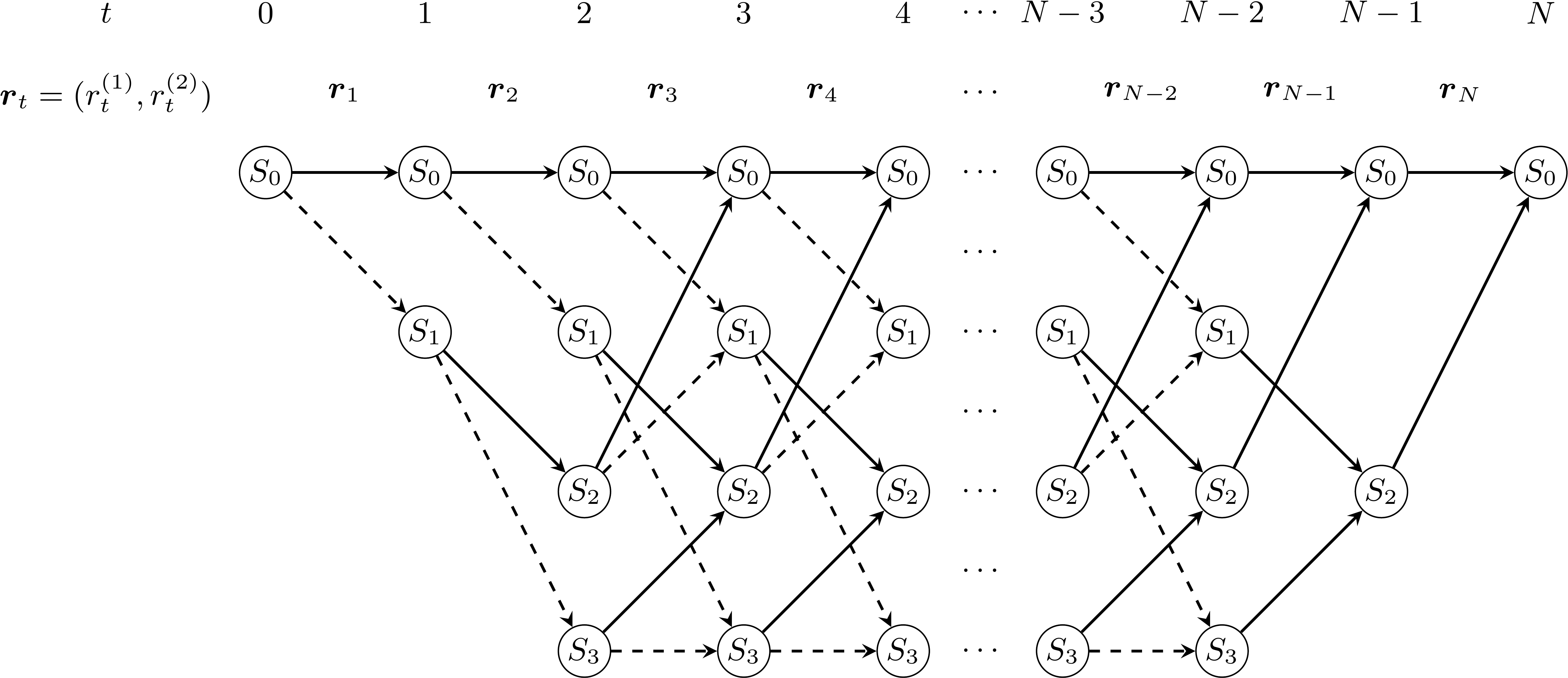}
  \caption{Trellis diagram for rate $1/2$, $K=3$ convlutional code. }
  \label{fig:trellis_cc}
\end{figure}

The coding scheme of the convolutional codes used in this study for the ISI MC
channel is same as the traditional conventional codes. In Fig.
\ref{fig:encoding}, a simple encoding structure of conventional codes $(7,5)$
with code rate $R=1/2$ and constraints length $K=3$ is presented, which can be
easily implemented in the nanomachines with limited computational capability.
Fig. \ref{fig:trellis_cc} is the trellis diagram of the conventional code
$(7,5)$.

\subsection{Decoding Scheme}
\label{sec:decoding}

The classical decoding method of conventional codes is Viterbi algorithm, a
maximal-likelihood decoding algorithm, which uses Hamming distance to measure
the distance between probable transmitted sequences and received sequences for
hard decision decoding and Euclidean distance for soft decision in the additional
white Gaussian noise (AWGN) channel. The main challenge in this study is that
the bit errors in the ISI MC channle is mainly
caused by the crossovers between the particles. In this section the crossover
distance proposed in Section \ref{sec:crossover-distance} is used as the metric
in Viterbi algorithm for the ISI MC channel. By minimizing 
crossover distance between the received sequence and the probable transmitted
sequence the maximal likelihood (ML) estimation of the tranmitted bits can be achieved.

In order to apply the crossover distance in Viterbi algorithm (VA) for
convolutional codes, there are two key issues need to be addressed:
\begin{enumerate}
\item The Viberbi algorithm finds the max-sum or max-product path in the
  trellis, where the metric of each path is addible and extends along with each
  symbol receiving sequentially. However the proposed crossover distance is
  derived from the crossover vector which is computed from the entire sequence
  blocks. Such that a sequential algorithm for computing crossover vector
  between the receveived sequence and the probable tranmitted sequences is
  developed to meet the requirement of Viterbi algorthm.
\item It is assumed that the numbers of ``0'' and ``1'' bit in the input and
  output sequences of the ISI MC channel are same, which may not be satisfied for
  the case of partial receiving and sequential processing in Viterbi algorithm.
  The delayed decision of the received symbols and the extension of the current
  probable path branches are applied to ensure this assumption.
\end{enumerate}

The modified Viterbi algorithm for the ISI molecular channel is presented in the
Algorithm \ref{alg:modViterbi}, where $S_0, S_1, \cdots, S_{M-1}$ is the
internal states of the convolutional decoder, $d_i$ is the current distance
metric associated with state $S_i$ and $R_i=\{S_{i,1}, S_{i,2},\cdots, S_{i,t}\}$
is the vector of the states associated with the optimal branch at current state
$S_i$. $M=2^{K-1}$ is the number of the internal states, $N$ is the length of
the transmitted sequence, $n$ is the number of the output bits corresponding to
each state transfer. For the convolutional codes in Fig. \ref{fig:encoding} and
Fig. \ref{fig:trellis_cc}, $K=3$, $M=4$ and $n=2$.

\begin{algorithm}
  \caption{Modified Viterbi algorithm} \label{alg:modViterbi}
  \begin{algorithmic}[1]
    \STATE $d_i=0, R_i=()$, $i=0,1,\cdots,M-1$
    \FOR{$t=1$ to $N$}
    % \STATE Receive new bits $(r^{(1)}_t, \cdots, r^{(2)}_t)$,
    \STATE Prefetch the received bits, $\vect{r}$
    \FOR{$i=0$ to $M-1$}
    \FORALL{$S_j$ such that $S_j$ transfers to $S_i$}
    \STATE Extending the transmitted bits  $\vect{x}_j$ along the branch in trellis,
    \STATE Update the input-ouput bijective function $L_j$ for $\vect{x}_j\to \vect{r}$
    \STATE $\Delta_j=0$
    \FOR{$k=(t-1)n$ to $tn-1$}
    \STATE Calculate $l_{L_j,k}$  according to \eqref{eq:clevel_k}
    \STATE $\Delta_j=\Delta_j+W_c(l_{L_j,k})$
    \ENDFOR
    \ENDFOR
    \STATE Update $d_i=\min_j (d_j+\Delta_j)$
    \STATE Update $R_i= R_j | j$, appending $j$ to state vector $R_j$.
    \ENDFOR
    \ENDFOR
    \STATE Recover the input bit sequence from $R_m$, where $m=\arg \min_id_i$
  \end{algorithmic}
\end{algorithm}

In Algorithm \ref{alg:modViterbi}, the input-output bijective function $L_j$ is
updated by extending the branch along the trellis path and prefetching the
received bits as long as the bijective mapping can be generated between the
current received bits and the encoded bits for state transfer $S_j\to S_i$.

The additivity of the crossover distance defined in \eqref{eq:DvWeight} ensures
that the distance mertic $d_i$ in Algorithm \ref{alg:modViterbi} can be
calculated sequentially and compared partially. Distance weight of level-$l$
crossover, $W_c(l)$ defined in \eqref{eq:Wc} corresponds to the probability of
the occrrence of $l$-level crossover, which is usually a float point number. In
order to simplify the computation of the crossover distance, \emph{level-sum
  distance} is defined as
\begin{align}
  \label{eq:level-sum}
  D_{\vect{v}}(\vect{x}, \vect{y})&=\sum_{k=1}^N l_k,
\end{align}
which is equialent to define $W_c(l)$ as
\begin{align} 
  W_c(l)=l&=\log P_c(0)-\log P_c(l) \nonumber \\
   &=\log \frac{P_c(0)}{P_c(l)}. \label{eq:Wc-level-sum}
\end{align}
That is, the probability of level-$l$ crossover is $\ee^{-l}P_c(0)$.
By choosing distance weight $W_c(l)$ defined in
\eqref{eq:Wc-level-sum}, the level-sum distance metric can be easily applied in
Algorithm \ref{alg:modViterbi}, which is easier to implement in the simple devices
without float-point number computation capability. The bit error rate (BER) performance loss of
level-sum distance based Viterbi algorthm is small compared with crossover
distance based algorithm, which is presented in Section
\ref{sec:numerical-results}.

In theory, Theorem \ref{thm:crossvect} proves that
the sum of crossover vector is minimized based on Algorithm \ref{alg:crossvect},
which also indicates that the survived branch in Algorithm \ref{alg:modViterbi} has the
minimized level-sum distance for the received bit sequence.

%The theoretic analysis of the BER performance of the proposed decoding scheme is
%still an open question, which can draw on the analysis of the tranditional
%Viterbi algorithm.

\section{Numerical Results}
\label{sec:numerical-results}

In this section the performance of the crossover-distance based decoding scheme
of convolutional codes is first discussed through simulation. And
theoretical approximations of BER is also provided in this section

\subsection{Simulation Results}
The bit error rate (BER) performance of proposed decoding
scheme is compared with various channel codes including (7,4) Hamming codes, ISI-free
codes \cite{shih_channelCodeMC_2013JSAC} and Hamming-distance based conventional
codes. Table \ref{tab:codes-list} is the conventional codes used in this study,
which are rate 1/2 maximum free distance codes listed in \cite{lin2001ECC}. In
Table \ref{tab:sim-parm}, the parameters of the molecular channel model in
numerical simulation are provided. 
\begin{table}
  \centering
  \caption{The rate 1/2 convolutional codes used in the simulation}
  \label{tab:codes-list}
  \begin{tabular}{|c|c|}
    \hline
    Constraint length $K$ & Generate in Octal \\ \hline
    3 & $[5,7]$  \\ \hline
%    4 & $[13,17]$ \\ \hline
    5 & $[27,31]$ \\ \hline
%    6 & $[53,75]$ \\ \hline
    7 & $[117,155]$ \\ \hline
  \end{tabular}  
\end{table}

\begin{table}
  \centering
  \caption{Simulation parameters}
  \label{tab:sim-parm}
  \begin{tabular}{|c|c|}
    \hline
    Temperature ($T_{\mathrm{K}}$)  & 298K \\ \hline
    Viscosity of water ($\eta$) & 0.894 mPa$\cdot$s \\ \hline
    Molecule radius ($r$) & 10 nm \\ \hline
    Distance of Tx and Rx ($d$) & 10 $\mathrm{\mu}$m \\ \hline
    Diffusion constant ($D$) & $2.44\times 10^{-11}\mathrm{m}^2/\mathrm{s}$ \\ \hline
    Drift velocity ($v$) & 10 $\mathrm{\mu}$m/s \\ \hline

  \end{tabular}
\end{table}

\begin{figure}
  \centering
  \includegraphics[width=0.45\textwidth]{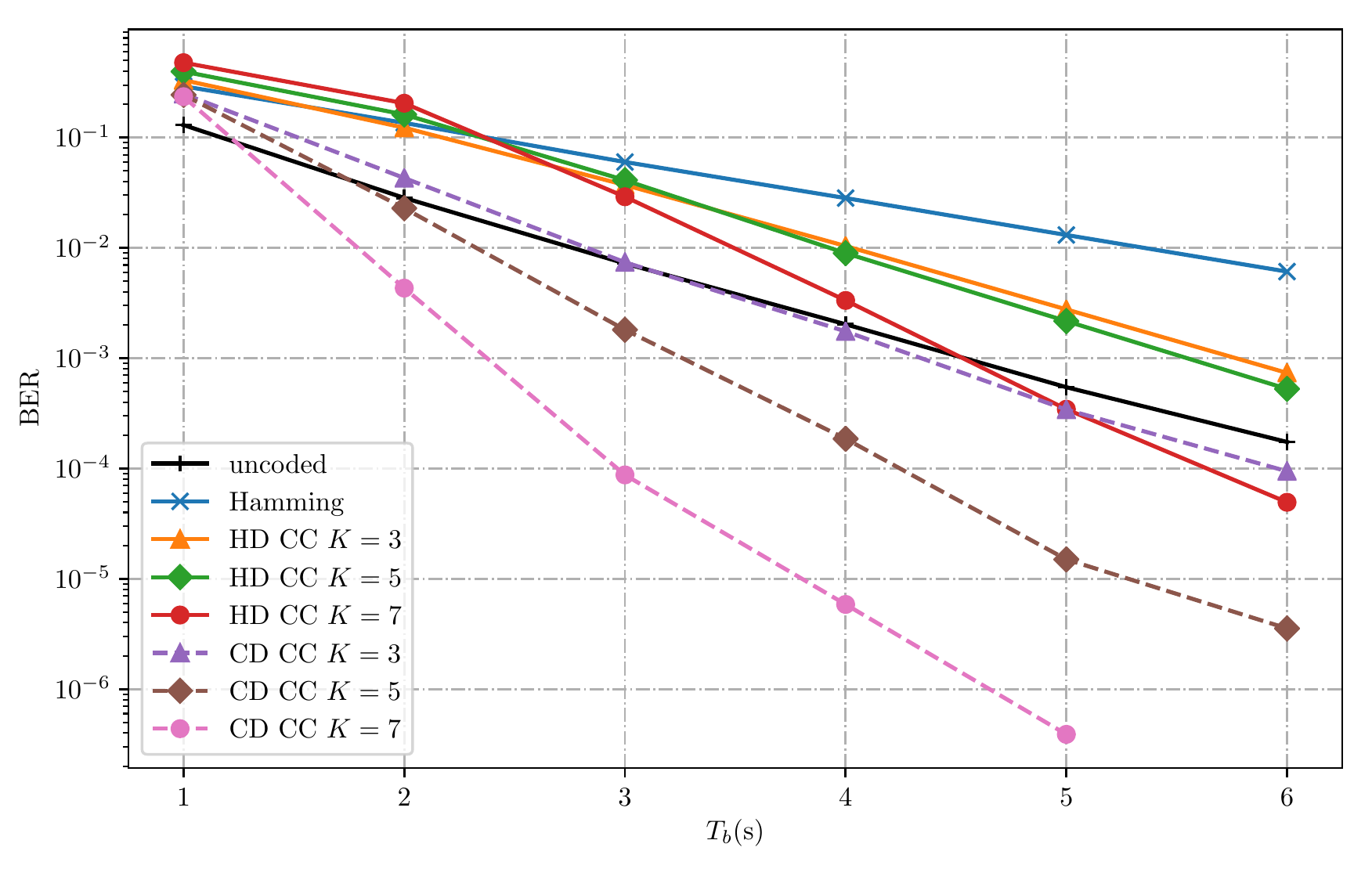}
  \caption{BER versus $T_b$ for uncoded system, Hamming code, Hamming-distance
    convolutional codes (HD CC) and Crossover-distance convolutional codes (CD CC).}
  \label{fig:ber-hamming-vs-crossover}
\end{figure}

In order to compare the performance of different channel codes fairly, the BER
of each codes is with the same information bit throughput. Define $T_b=T_s/R$
as the equivalent information bit interval, where $R$ is the code rate. Fig.
\ref{fig:ber-hamming-vs-crossover} shows the BER performance versus the
equivalent information bits interval. As illustrated in Fig.
\ref{fig:ber-hamming-vs-crossover}, the BER performance of Hamming-distance
based channel codes including Hamming codes, convolutional codes with constraint
length $K=3,5,7$ is not better than uncoded system significantly, such that the
redundant bits introduced by these codes can not correct errors in the ISI
MC channel effectively. Whereas the crossover-distance based
convolutional codes can cope well with the errors caused by ISI in MC
channel, which indicates that the proposed crossover distance is a suitable
tool for studying channel codes in the diffused-based molecular communication channel.

\begin{figure}
  \centering
  \includegraphics[width=0.45\textwidth]{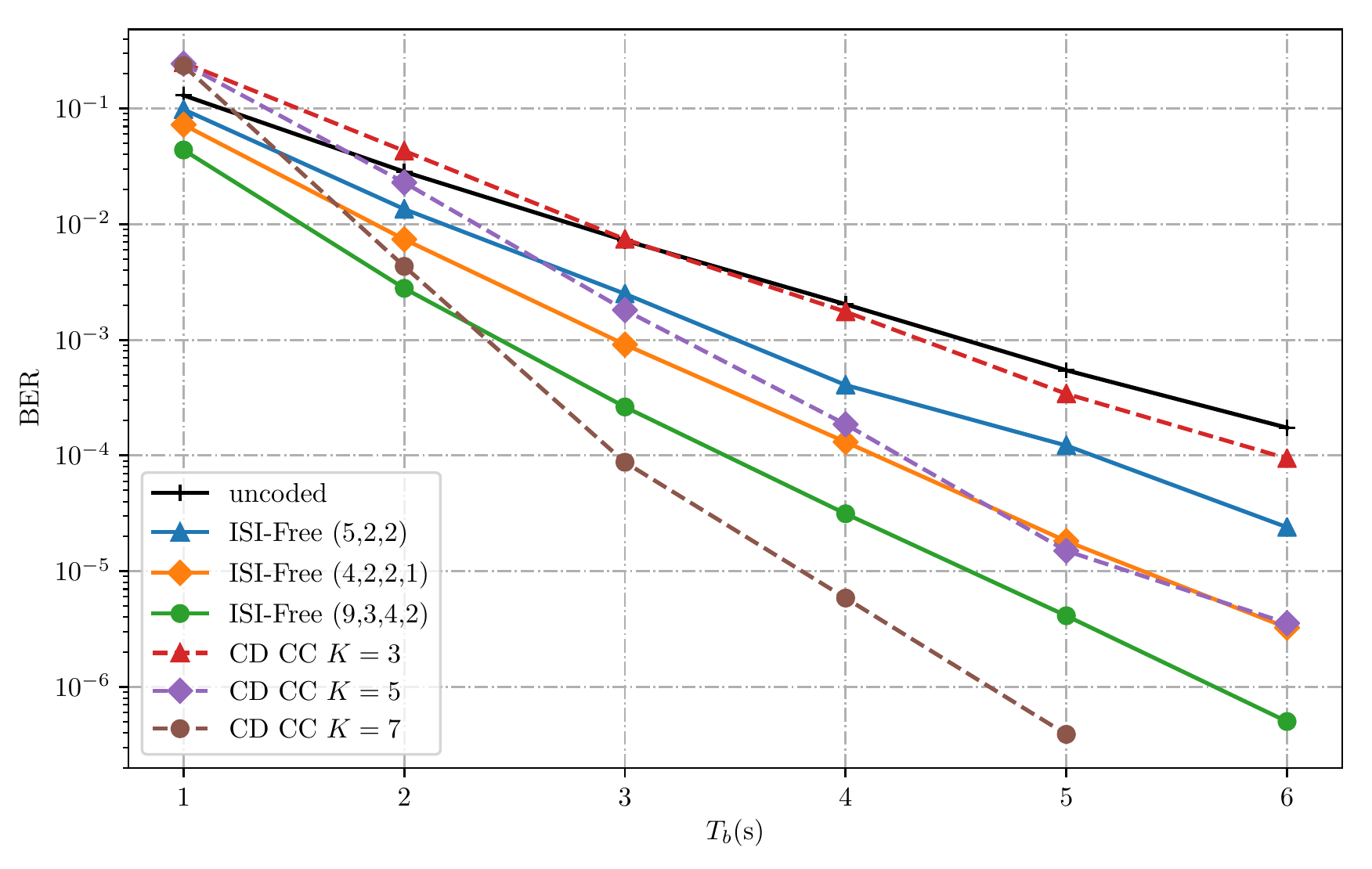}
  \caption{BER versus $T_b$ for uncoded system, Crossover-distance convolutional
    codes (CD CC) and ISI-Free codes.}
  \label{fig:ber-crossover-vs-isi-free}
\end{figure}

The ISI-free codes proposed in \cite{shih_channelCodeMC_2013JSAC} is a new code
family for diffusion based MC channels with good BER performance compared to
uncodes transmission under the same throughputs. In Fig.
\ref{fig:ber-crossover-vs-isi-free}, the crossover-distance based convolutional
codes is compared with the three kinds of ISI-free codes. It is shown that the
proposed convolutional codes with constraints length $K=5, 7$ can achieve the
similar or better BER performance as ISI-free codes under the same throughputs. It
should be noted that as a kind of short-length block code ISI-free codes have
lower decoding complexity than the convlutional codes. On the other hand, as a
ML-decoding channel codes, crossover-distance based convolutional codes can
easily achieve better performance by utilizing the classical research resources on
convolutional codes and its expansion such as Turbo codes, which is an important
advantage in some reliability-sensitive scenario when the transmitter is simple
nanodevice and the receiver is traditional device with powerful computation
capability.

\begin{figure}
  \centering
  \includegraphics[width=0.45\textwidth]{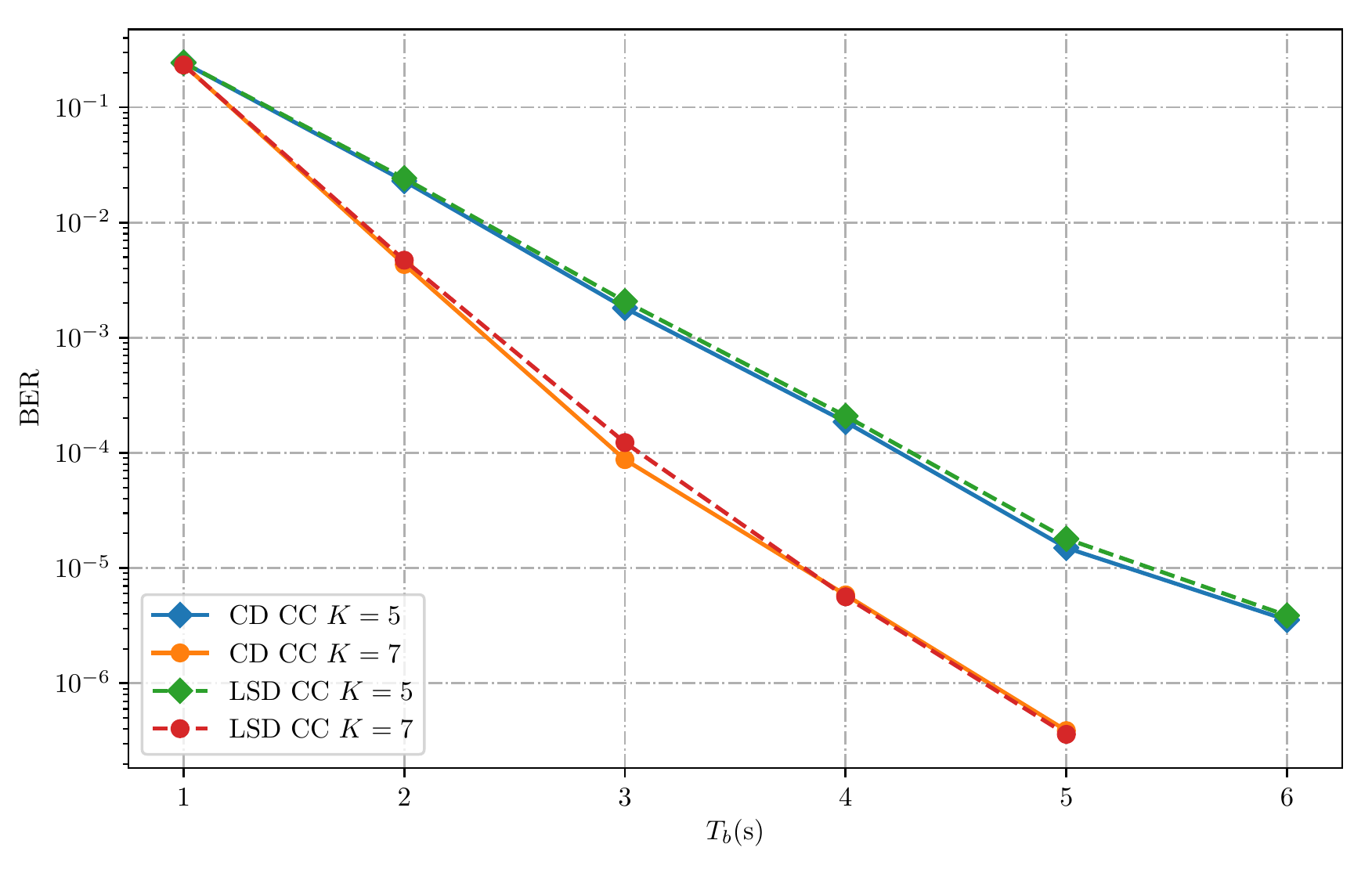}
  \caption{BER versus $T_b$ for Crossover-distance convolutional
    codes (CD CC) and Level-sum-distance (LSD CC) convolutional codes}
  \label{fig:ber-crossover-vs-level-sum}
\end{figure}

In Section \ref{sec:decoding-scheme}, level-sum distance is proposed as a
simplified edition of the crossover-distance. The numerical results in Fig.
\ref{fig:ber-crossover-vs-level-sum} shows that the BER performance loss caused
by inaccurate estimation of crossover distance through level-sum distance is
trivial, which facilitates the use of proposed decoding scheme in engineering.


\subsection{Approximations of the BER}
In order to approximate the BER of the proposed crossover-distance convolutional
codes, we use codeword input-output weight enumerating function (IOWEF) \cite{lin2001ECC} to
enumate each error paths on the trellis graph of the convolutional codes. For
each sending codeword $\vect{x}$, the corresponding $n_{\vect{x}}$ error decoding
codewords are $\vect{R}^{(1)}, \vect{R}^{(2)},\cdots, \vect{R}^{(n_\vect{x})}$.
Suppose the $\vect{x}$ is transformed to the received codeword $\vect{y}_j$ by the
crossover vector $\vect{v}_j$, then the possibility of the decoding error event
occurring for $\vect{x} \to \vect{R}^{(k)}$ is
\begin{align}
  \label{eq:PRk}
  P_{\vect{R^{(k)}}|\vect{x}}&=\sum_{j} P(\vect{y}_j|\vect{x})
                        \I\{D(\vect{x}, \vect{y}_j)>D(\vect{R}^{(k)}, \vect{y}_j)\} \\
                      &= \sum_{j}P(\vect{v}_j)\I\{D(\vect{x}, \vect{y}_j)>D(\vect{R}^{(k)}, \vect{y}_j)\}
\end{align}

\begin{figure}
  \centering
  \includegraphics[width=0.45\textwidth]{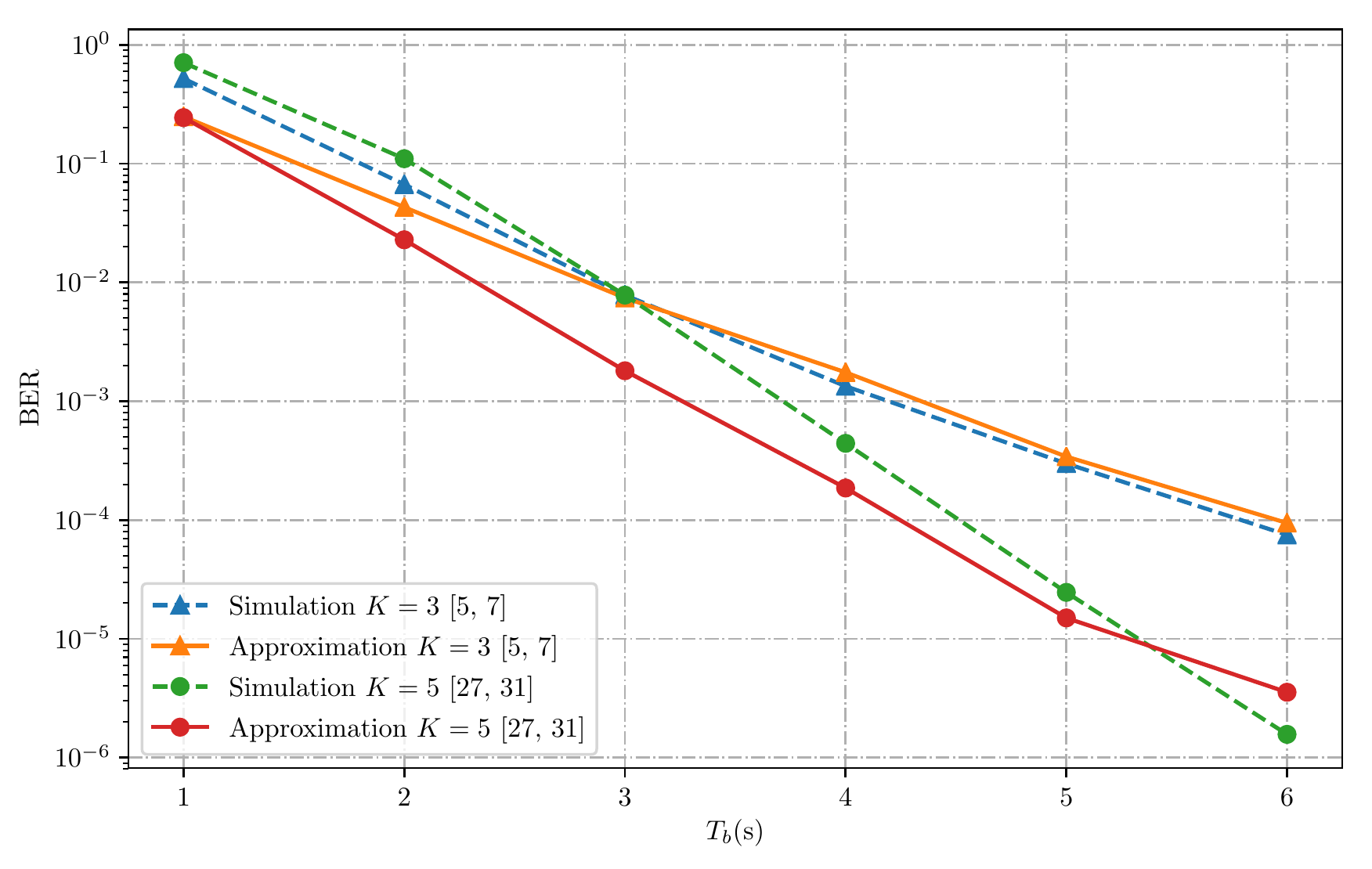}
  \caption{BER Comparison between the approximation and simulation of
    Crossover-distance convolutional codes}
  \label{fig:ber-appoxy-vs-simulation}
\end{figure}

Then the error bit rate can be approximated as
\begin{align}
  \label{eq:appox-ber}
  P_b\approx \sum_{\vect{x}}P_{\vect{x}}\sum_{k=1}^{n_{\vect{x}}} W_{\vect{x}, \vect{R}^{k}}P_{\vect{R^{(k)}}|\vect{x}}
\end{align}
where $W_{\vect{x}, \vect{R}^{(k)}}$ is the number of different information bits
between codeword $\vect{x}$ and corresponding error decoding codeward
$\vect{R}^{(k)}$. Fig. \ref{fig:ber-appoxy-vs-simulation} is the BER
comparison between the approximation and simulation of Crossover-distance
convolutional codes. It is shown that formula \eqref{eq:appox-ber} can be
used to approximate the BER performance apparently.

\section{Conclusion}
\label{sec:conclusion}

In this paper, crossover distance is proposed for the decoding of convolutional
codes in the ISI molecular communication channel, which can be estimated through
the minimum crossover vector between tranmitted bit sequence and received bit
sequence. We proved that the minimum crossover vector has the minimum
crossover-level sum and maximum. The algorithm with a time complexity $O(n^2)$
is also presented to calculate the minimum crossover vector. By using crossover
distance, a modified Viterbi decoding scheme is given for decoding convolutional
codes in the ISI molecular communication channel. The numerical results on BER
performance of various channel codes show that the proposed decoding scheme enhances
the reliability of the diffusion based molecular communication significantly. In
addition, the theoretical approximation of the BER for decoding convolutional
codes based on crossover-distance is aslo discussed briefly.

The effort in this paper also provides an integration between the traditional
channel codes theory and the novel molecular communication. It is implied that the
proposed concepts including crossover vector and crossover distance maybe capture the
characteristics of the ISI MC channel, which can be used as a tools to
exploit the new channel codes for molecular communication channel. Moreover, the
proposed decoding scheme can also be modified to suit other molecular modulation,
which is interesting and left for future work.

%\IEEEPARstart{T}{his} demo file is intended to serve as a ``starter file''
%for IEEE Communications Society journal papers produced under \LaTeX\ using
%IEEEtran.cls version 1.8b and later.
% You must have at least 2 lines in the paragraph with the drop letter
% (should never be an issue)

% needed in second column of first page if using \IEEEpubid
%\IEEEpubidadjcol

\section*{Acknowledgment}
We thank Prof. Dongning Guo, Northwestern University, for the discuss on this
study and for the comments that greately improved the manuscript.

\bibliographystyle{IEEEtran}
\bibliography{IEEEabrv,isi-ccc}

% biography section
% 
% If you have an EPS/PDF photo (graphicx package needed) extra braces are
% needed around the contents of the optional argument to biography to prevent
% the LaTeX parser from getting confused when it sees the complicated
% \includegraphics command within an optional argument. (You could create
% your own custom macro containing the \includegraphics command to make things
% simpler here.)
%\begin{IEEEbiography}[{\includegraphics[width=1in,height=1.25in,clip,keepaspectratio]{mshell}}]{Michael Shell}
% or if you just want to reserve a space for a photo:

% insert where needed to balance the two columns on the last page with
% biographies
%\newpage

% You can push biographies down or up by placing
% a \vfill before or after them. The appropriate
% use of \vfill depends on what kind of text is
% on the last page and whether or not the columns
% are being equalized.

%\vfill

% Can be used to pull up biographies so that the bottom of the last one
% is flush with the other column.
%\enlargethispage{-5in}

% that's all folks
\end{document}